\documentclass[superscriptaddress,twocolumn,showpacs,pra]{revtex4-1}

\usepackage{graphicx} 
\usepackage{amssymb}

\begin{document} 

\title{Robust Mesoscopic Superposition of Strongly Correlated Ultracold Atoms} 

\author{David W. Hallwood}
\affiliation{New Zealand Institute for Advanced Study and Centre for Theoretical Chemistry and Physics, Massey University, Private Bag 102904, North Shore, Auckland 0745, New Zealand.} 
\affiliation{School of Physics and Astronomy, University of Leeds, Leeds LS2 9JT, United Kingdom.} 
\author{Thomas Ernst}
\affiliation{New Zealand Institute for Advanced Study and Centre for Theoretical Chemistry and Physics, Massey University, Private Bag 102904, North Shore, Auckland 0745, New Zealand.} 
\author{Joachim Brand}
\affiliation{New Zealand Institute for Advanced Study and Centre for Theoretical Chemistry and Physics, Massey University, Private Bag 102904, North Shore, Auckland 0745, New Zealand.} 
\email{J.Brand@massey.ac.nz}
\date{\today}

\pacs{03.75.Gg,67.85.Hj,37.25.+k,03.67.Bg}

\begin{abstract}
We propose a scheme to create coherent superpositions of annular flow of strongly-interacting bosonic atoms in a 1D ring trap. The non-rotating ground state is coupled to a vortex state with mesoscopic angular momentum by means of a narrow potential barrier
and an applied phase that originates from either rotation or a synthetic magnetic field.
We show that superposition states in the Tonks-Girardeau regime are robust against single-particle loss due to the effects of strong correlations. The coupling between the mesoscopically distinct states scales much more favorably with particle number than in schemes relying on weak interactions, thus making particle numbers of hundreds or thousands feasible. 
Coherent oscillations induced by time variation of parameters may
serve as a `smoking gun' signature for detecting superposition states. 
\end{abstract}

\maketitle 

\section{Introduction}
Quantum superpositions of macroscopically distinct states are important for our understanding of quantum mechanics~\cite{leggett_85} and carry great promise for enhanced precision measurement techniques~\cite{giovannetti_04}. 
A conceptually simple example is the superposition $|N,0\rangle + |0,N\rangle$, where all $N$  particles occupy either one or the other of two accessible modes (e.g. spin orientations). Due to their inherent fragility, such maximally entangled {\em NOON} states~\cite{note1} engineered in optics and spin systems have been limited to 10 particles~\cite{jones_09}. 
The related mesoscopic superpositions of flux states consisting of ${10^9}$ Cooper pairs observed in superconducting rings~\cite{wal_00,friedman_00} have proven more robust but their microscopic nature is debated~\cite{leggett_87,korsbakken_07,marquardt_08,korsbakken_09}.
Proposals to create mesoscopic superpositions of ultra-cold atoms have so far entirely focused on NOON and closely related states~\cite{optical,bader_09,weiss,streltsov} and have not yet been realized. Such proposals
suffer from severe limitations due to decoherence~\cite{zurek_03} and the unfavorable scaling of precision and time scales needed to produce these states~\cite{hallwood_07,frazer_07}.

In this paper we propose a simple and experimentally accessible scheme for producing large, robust quantum superposition states of ultra-cold atoms. NOON states are so fragile because the loss of a single constituent particle, if discriminant between either of the two available modes, destroys the superposition. We show that many-particle superposition states can be made more robust by making use of interactions between the atoms. The intuition is that correlations due to particle interactions spread single-particle observables over many modes. This disguises the origin of any lost particle and allows the superposition to survive. In addition, strong interactions remove degeneracies and therefore decrease sensitivity to environmental fluctuations.

\begin{figure}
\includegraphics[width=8cm]{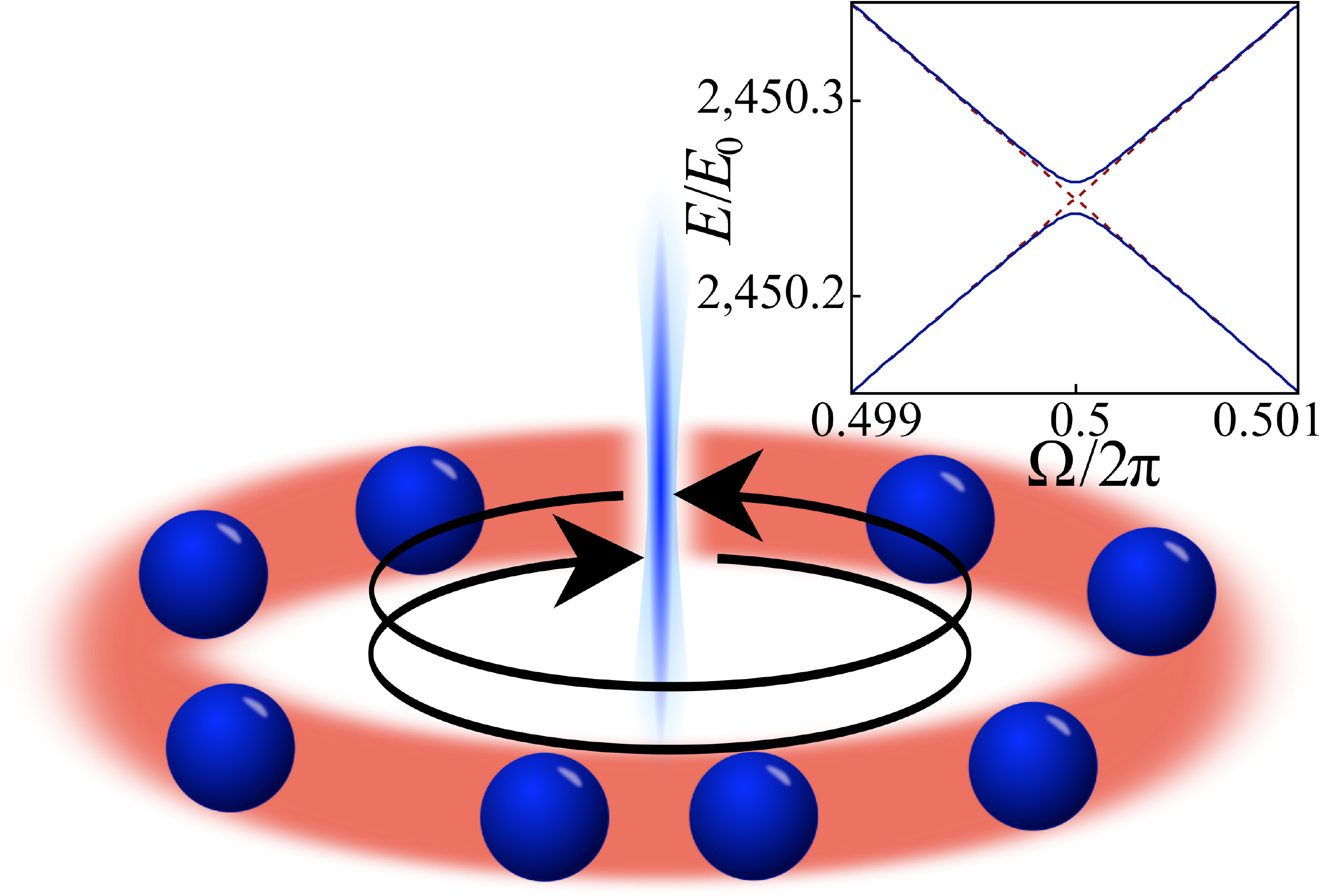}
\caption{ (Color online)
Strongly-correlated atoms trapped in a narrow ring with a rotating barrier. The inset shows the energy levels of the  super-current states with total angular momentum $K=0$  and $K=N \hbar$, respectively, as a function of the rotational phase $\Omega$ (dashed lines), and the lowest energy levels of $N=99$ atoms in the Tonks-Girardeau regime with barrier strength $b/L = 0.008 E_0$ (full lines).
}
\label{fig:system}
\end{figure}

\section{The system}
We apply this idea to bosonic atoms confined to a thin ring-shaped trap \cite{gupta_95,ryu_07} that is intersected by a  focused blue-detuned laser beam, which creates a potential barrier for the atoms
as illustrated in Fig.~\ref{fig:system}.
One-dimensional Bose gases with variable repulsive interactions have already been realized in linear traps where the interactions were tuned by means of a Feshbach resonance or by adjusting the trap geometry \cite{kinoshita_04,haller_09}. 
We model the system of $N$ atoms of mass $M$ in a loop of circumference $L$  by the one-dimensional Hamiltonian
\begin{eqnarray}
H \! = \! \sum_{i=1}^N \left[ \frac{\hbar^2}{2M}\!\!\left( \!-i\frac{\partial}{\partial x_i} \!- \!\frac{\Omega}{L} \right)^2 \!\!\!+\!b \delta(x_i) \!+\! {g} \sum_{i<j}^N \delta(x_i\!\!-\!\!x_j) \right] ,
\label{eq:ham1}
\end{eqnarray}
where $x=\theta L/2\pi$ is the atom's position on the circumference of the loop and ${g}$ is the effective one-dimensional interaction strength between the atoms~\cite{olshanii_98}. The smallest nonzero kinetic energy of a single particle
\begin{equation}
 E_0 = 2 \pi^2 \hbar^2/(M L^2)
\end{equation}
provides a natural unit of energy for this system.
The narrow barrier with strength $b$ is rotated with constant tangential velocity  $v=\hbar\Omega/(ML)$ along the circumference of the ring and the Hamiltonian~(\ref{eq:ham1}) is formulated in the co-rotating frame of reference. Alternatively, the rotational phase $\Omega$ can be applied in a non-rotating  system  by means of a synthetic magnetic field as recently demonstrated at NIST \cite{Spielman}. 

In the absence of the barrier ($b=0$), the angular momentum in the ground state is quantized to integer multiples of $N \hbar$~\cite{bloch_73,cherny_09}. In this case, the Hamiltonian (\ref{eq:ham1}) describes the integrable Lieb-Liniger model~\cite{lieb_63} with energies of different angular momentum states shifted with respect to each other by a Galilean transformation due to the rotational phase $\Omega$.
A finite barrier ($b>0$) couples  
states with different total angular momentum, leading to the avoided level crossing seen in the inset in Fig.~\ref{fig:system}. At the precise position of the avoided crossing, an effective two-level system is realized with eigenstates being 50/50 superpositions of the two angular momentum states. By adiabatically changing the applied phase from zero into the avoided crossing the superposition is created, while a rapid, non-adiabatic procedure leads to coherent oscillations between the two states. The oscillations with the period of $\Delta E/ \hbar$, where $\Delta E$ is the level splitting at the avoided crossing
\cite{nunnenkamp_08}, can be used to detect the presence of superposition states. The quantized total angular momentum is measured by dropping the trap and observing the interference pattern develop a hole ($K=N\hbar$) or a peak ($K=0$) along the ring axis \cite{ryu_07}. 

In order to observe coherent oscillations, the level splitting $\Delta E$ should be larger than the rate of decoherence. We have calculated how $\Delta E$ scales with particle number in different regimes of interaction strength. 
While weak interactions lead to NOON states, the unfavorable scaling of $\Delta E$ and sensitivity to particle loss severely limit the attainable particle numbers. In the strongly-interacting regime, however, $\Delta E$ grows with particle number and the superposition states become robust against single particle loss.

\section{Numerical Simulation}
\begin{figure}
\includegraphics[width=8cm]{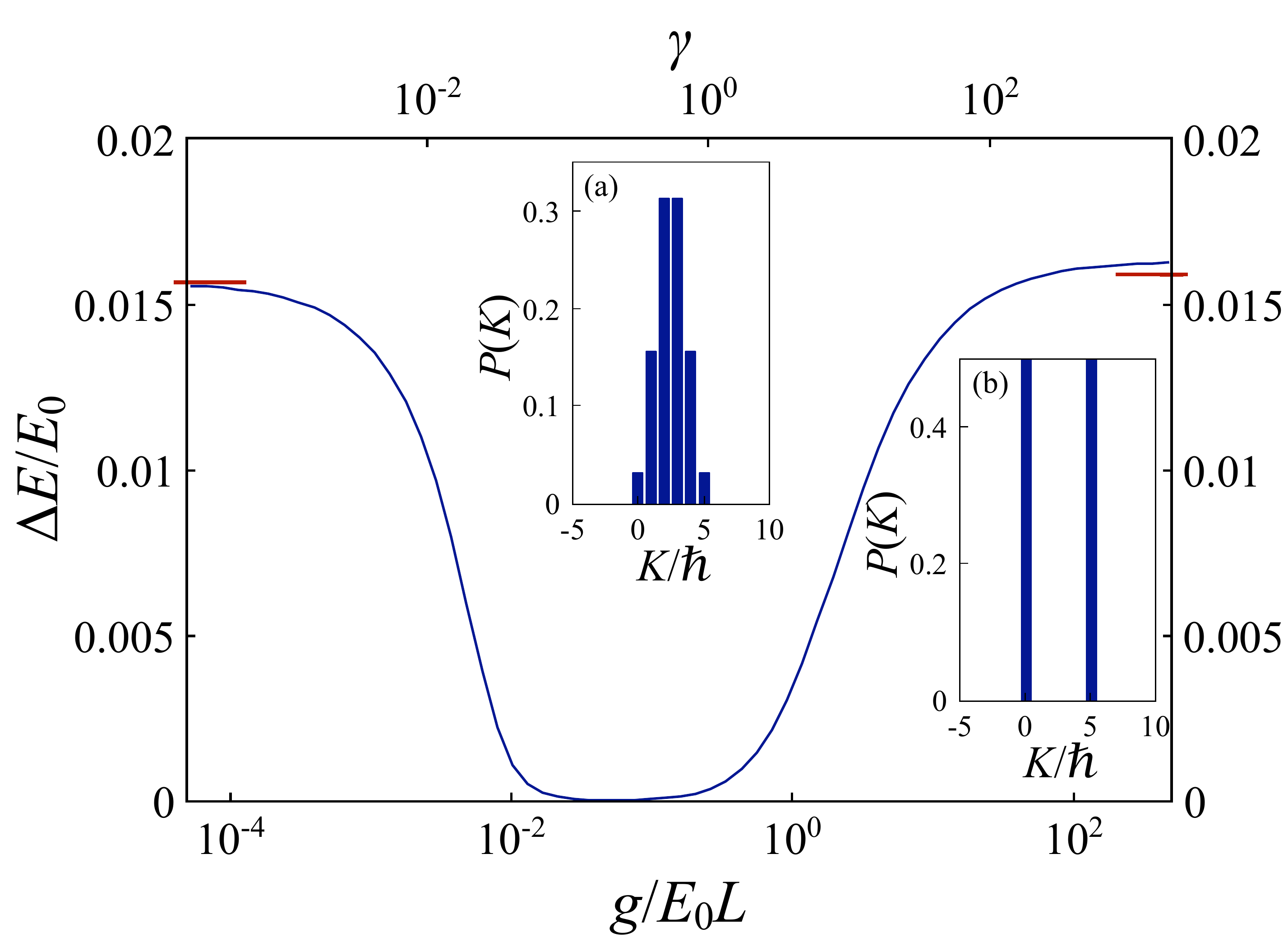}
\caption{ (Color online) 
The level splitting $\Delta E$ between the ground and first
  excited state at $\Omega=\pi$ is plotted over interaction strength ${g}$ for 5 atoms with $b/L =
  0.008E_0$. The top horizontal axis shows the Lieb-Liniger parameter $\gamma={g} 2\pi^2/(E_0 L N)$. The dashes on the figure margins indicate analytic results for non-interacting ($g=0$) and strongly interacting atoms ($g=\infty$). 
Subplots show the distribution of the total angular momentum for (a) $g=0$
and (b) $g=\infty$.
}
\label{fig:DEvg}
\end{figure}

We have simulated the system for $N=5$ particles by numerical 
diagonalization of the Hamiltonian (\ref{eq:ham1}).
In order to obtain accurate results into the strongly-interacting regime,
we have developed an effective Hamiltonian for the one-dimensional Bose gas \cite{Suzuki}.
In Appendix~\ref{ap:numerics} we show that a rescaled interaction strength  $\tilde{g}=g(1+g/g_0)^{-1}$
produces exact energies and eigenstate projections for two particles when the Hamiltonian is represented in a truncated occupation number basis from $r$ angular momentum modes with $g_0\approx rLE_0/2$, where we have ignored small energy dependent
terms. 
For larger particle numbers we use the same value of $\tilde{g}$ and monitor the error by comparing with the exact Lieb-Liniger result. We still find a significant improvement compared to the unscaled Hamiltonian. We have used $r=20$ modes for all calculations (except Fig.~\ref{fig:loss}a).
In the Tonk-Girardeau regime, where the error is the largest, the relative error after rescaling is less than 3\%, which is a factor of 8 times smaller than without the rescaling. In addition to the level splitting $\Delta E$ shown in Fig.~\ref{fig:DEvg}, the simulations also provide insight into the nature and the robustness properties of the superposition states as shown in Fig.~\ref{fig:loss}. 

\section{Non-Interacting Atoms}
Before discussing the 
many-particle states in different interacting regimes we first consider
the solutions for a single atom in the system (see Appendix~\ref{ap:single_particle}). These are plane waves with angular momentum $n\hbar$ in the absence of the barrier. At $\Omega=\pi$ the pairs of energy levels with $k_1= -n \hbar$ and $k_2 = (n+1)\hbar$ are degenerate  but this degeneracy is lifted for non-zero barrier strength $b$. 
Expressions for the energy levels $\varepsilon_\mu$ and eigenstates are obtained  using the transformation
$\Psi(x) =\phi(x)e^{i\Omega x/L}$, where $\phi(x)=\phi(x+L)e^{i\Omega}$ and employing the ansatz $\phi(x)=e^{i2\pi \alpha x/L} + A e^{-i2\pi \alpha x/L}$. 

For the case of $\Omega = \pi$, the discrete solutions $\alpha_{\mu}$ with $\mu = 0,1,2,...$ correspond to the single particle energy levels $\varepsilon_{\mu} = \alpha_{\mu}^2 E_0$, and are the roots of $2\pi \hbar^2\alpha_{\mu}/MLb = -\tan(\pi \alpha_{\mu})$ for odd $\mu$ and $\alpha_{\mu} = (\mu+1)/2$ for even $\mu$.  The $\nu$th avoided crossing has a level splitting of $\varepsilon_{2\nu-1}-\varepsilon_{2(\nu-1)} \sim b/L$ for small barrier $b\ll 2 \nu E_0 L$ but reaches a constant value $\varepsilon_{2\nu-1}-\varepsilon_{2(\nu-1)} \sim (\nu-1/4) E_0$ for large impenetrable barrier $b\gtrsim 2\nu E_0 L$.

For non-interacting atoms the energy gap $\Delta E$ is the energy
required to excite a single atom from the ground state across the
first avoided level crossing (see dash on the far left of
Fig.~\ref{fig:DEvg}). Instead of a binary superposition, however, we
find a binomial distribution involving many different momenta, as seen in Fig.~\ref{fig:DEvg}a.

\section{Tonks-Girardeau Regime}
The situation becomes more interesting when repulsive interactions remove the near degeneracies of all angular momentum eigenstates but the $K=0$ and the $K=N\hbar$ states. The interacting quantum system is generally very difficult to model and our numerical approach is limited to a small number of particles. 
One exception is the Tonks-Girardeau limit of strong interactions and low densities where $\gamma \equiv gML/(\hbar^2 N) \gg 1$. Here, the bosons are strongly correlated as they cannot pass each other and undergo fermionization. The energy spectrum of the bosons in this limit is identical to that of non-interacting spinless fermions~\cite{girardeau_60} but the single-particle momentum distribution is different and spreads $\propto1/\sqrt{|k|}$ over an infinite range even at zero temperature~\cite{Lenard}.
Experiments have already validated the features of fermionization of ultra-cold atomic gases~\cite{paredes_04,kinoshita_04}.

A calculation of the energy gap in the Tonks-Girardeau limit proceeds as for non-interacting spinless fermions, using the single-particle results. For simplicity we assume an odd number of particles $N$. The ground state energy is found by summing the lowest $N$ single-particle levels, while the first excited state energy is found by substituting the highest level in the sum by the next higher level. The gap is given by the difference between these two energies $\Delta E = \varepsilon_N-\varepsilon_{N-1}$ and is marked on the far right of Fig.~\ref{fig:DEvg}. In the weak barrier regime determined by $b \ll N E_0 L$, we find that 
\begin{equation}
\Delta E = b/L + {\cal O}(b^2)
\end{equation}
scales {\em independently of particle number} while the maximum attainable gap energy is 
\begin{equation}
\Delta E_{\mbox{\tiny{max}}}=(N/2+1/4) E_0 .
\end{equation}
For a weak barrier and finite but strong interactions, the gap energy shows power law scaling with system size at constant density $\Delta E/(N E_0) \sim N^\alpha$ with $\alpha = -4/\gamma + {\cal O}(\gamma^{-2})$ according to to Luttinger liquid theory~\cite{svist96,cherny_09-1}, where $N E_0$ is the energy of the first supercurrent excitation.

\begin{figure}
\includegraphics[width=8cm]{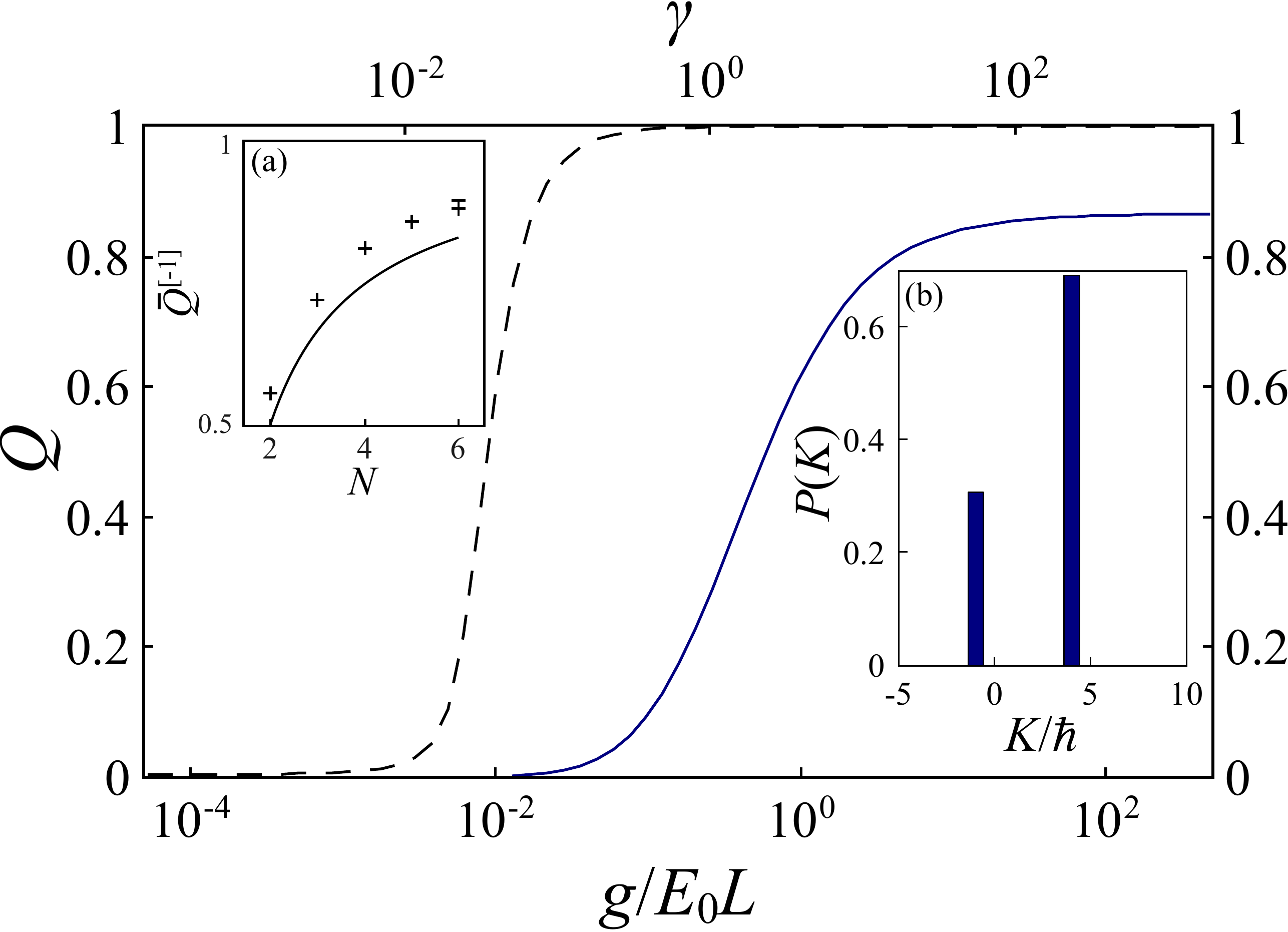}
\caption{ (Color online)
The quality of the superposition before and after removal of an atom. The dashed line shows the superposition quality $Q = 4{P(0)P(N\hbar)}$ for the ground state of the rotating system with parameters as in Fig.~\ref{fig:DEvg}. The solid line shows the average superposition quality $\bar{Q}^{[-1]}$ after the loss of an atom as described in the text. The crosses in inset (a) show $\bar{Q}^{[-1]}$ vs.\ atom number in the Tonks-Girardeau regime, and the solid line shows the result $1-1/N$ for non-interacting fermions for comparison.
The horizontal bar visible for $N=6$ estimates the error in the numerical result from an insufficient number $r$ of single particle modes by extrapolating from $r=14$ (the reported result) to larger $r$ by adding twice the difference from the $r=12$ result. All other results use $r=20$.
Inset (b) shows the total angular momentum distribution after removal of an atom with $k=1\hbar$ in the Tonks-Girardeau regime.}
\label{fig:loss}
\end{figure}

The distribution of the total angular momentum
in the ground state is shown in the two subplots in
Fig.~\ref{fig:DEvg} for (a) non-interacting atoms and (b) the
Tonks-Girardeau regime. The probability of the system having an
angular momentum of $K$, the neutral atom equivalent of flux or
current, is given by $P(K)=Tr(T_K \rho)$. Here $T_K=\sum_n |K,n\rangle
\langle K,n|$ is the projector onto states with total angular momentum
$K$, where the sum runs over all states with 	angular momentum $K$,
and $\rho = |\psi\rangle\langle \psi| / \langle \psi|\psi\rangle$ is
the density operator. Subplot (b) clearly shows a superposition of two
total angular momentum states, where, for $N$ atoms, the total
momentum difference of the two states is $N \hbar$. We quantify the
quality of a superposition between states with angular momentum $K_1$
and $K_2$ by 
\begin{equation}
Q=4{P(K_1) P(K_2)} . 
\end{equation}
As desired, the superposition quality  $Q$ varies between the maximum value of $1$ for an equal superposition and $0$ for angular momentum eigenstates. In Fig.~\ref{fig:loss} the dashed line shows $Q$ for the ground state (with $K_1=0$ and $K_2=N\hbar$). We see that $Q$ is very close to one beyond the minimum of $\Delta E$ at $g\approx 0.1 E_0 L$, where we also find that the distribution of total angular momentum does not change any more. 
The ground state is thus a binary superposition of total angular momentum for sufficiently large $g$.

\section{NOON State Regime}
The final feature of the $\Delta E$ curve in Fig.~\ref{fig:DEvg} that must be explained is the minimum around $g\approx 0.1 E_0 L$.
This is a regime where interactions are too weak to strongly correlate the atoms but strong enough to 
remove the degeneracies between states with intermediate angular momentum. We can assume the ground state for $\Omega = \pi$ to take the NOON form $|N,0\rangle + |0,N\rangle$, where the two numbers in the ket denote the occupation number of the single-particle modes with angular momentum $0$ and $\hbar$. For this to happen, the coupling to other angular momentum modes must be negligible. As shown in Appendix~\ref{ap:noon} this gives the conditions 
\begin{equation}
b\sqrt{N}/L \ll gN/2L \ll E_0 , 
\end{equation}
where the mean interaction energy per particle is small compared to the kinetic energy per particle and large compared to  the product of barrier energy and the square root of particle number.

The energy splitting  $\Delta E$ can now be calculated under the assumption that particles only access the two modes with $k=0$ and $k=\hbar$. The Schr\"odinger equation is solved by eliminating  states with intermediate angular momentum to yield (see Appendix~\ref{ap:noon}),
\begin{equation}
\Delta E = \frac{b^{N}}{g^{N-1}} \frac{2}{L}\frac{N}{(N-1)!} .
\label{eq:V}
\end{equation}
We find that $\Delta E$ is small and decreases faster than exponentially with atom number $N$. This makes the NOON state experimentally difficult to prepare through tuning $\Omega$ due to the long coherence times required~\cite{hallwood_07}. This difficulty can be overcome by creating a superposition in the strongly-correlated regime first and then adiabatically reducing $g$. The time to reach the NOON state is then no longer limited by $\Delta E$. The many-body dynamics involved in this process is left for future study.

\section{Particle Loss}
Important limiting processes for the lifetime of ultra-cold atom
experiments involve the loss of particles from the trap, e.g., due to
collisions with high-energy atoms from the background gas. When the
environment gains information about the state of the system this may
lead to the collapse of the superposition. We consider the most
detrimental  case of information about the angular momentum of the
atom being gained by the environment. The resulting state after
complete removal of a single atom with angular momentum $k\hbar$ from
the $N$ atom ground state $|\Psi\rangle$ is modeled by
\begin{equation}
|\Psi^{[-1]}_{k}\rangle = a_k
|\Psi\rangle/\sqrt{\langle\Psi|a_k^\dagger a_k|\Psi\rangle} . 
\end{equation}
We measure the robustness of the superposition state under particle loss
by the averaged quality 
\begin{equation}
\bar{Q}^{[-1]}= \sum_k Q^{[-1]}_k n_k/N ,
\end{equation}
where 
\begin{equation}
Q^{[-1]}_k = 4{P^{[-1]}(-k\hbar) P^{[-1]}((N-k)\hbar)}
\end{equation} 
is averaged over the angular momentum $k\hbar$ of the lost particle
weighted by the probability of finding an atom in mode $k$ and $n_k =
\langle \Psi^{[-1]}_k|a_k^\dagger
a_k|\Psi^{[-1]}_k\rangle$. Figure~\ref{fig:loss} shows
$\bar{Q}^{[-1]}$ after single-particle loss alongside the $Q$ for the
ground state as a function of the interaction strength $g$. The regime
around $g\approx 0.1 E_0 L$ features a high superposition quality but poor robustness against single-particle loss as expected for NOON states. In the Tonks-Girardeau regime, however, robustness increases dramatically and $\bar{Q}^{[-1]}$ is still of the order of one! Inset (b) in Fig.~\ref{fig:loss} demonstrates that the binary superposition is maintained even after removal of an atom, while inset (a) shows how robustness increases with atom number. 
The increased robustness is consistent with the spread-out nature of the single-particle momentum distribution. Measuring a single-particle's momentum is not sufficient to determine the total angular momentum of the state.

\section{Conclusion}
We have presented a scheme for producing a robust binary superposition with strongly correlated atoms in the ground state of a rotating system. A barrier couples two states that differ by angular momentum $N\hbar$ where $N$ is the number of atoms and the maximally achievable level splitting is proportional to $N$. Due to this favorable scaling, mesoscopic superpositions involving hundreds or thousands of atoms become feasible. Loading, e.g., 100 atoms of $^7$Li into a ring trap with radius $50\mu$m leads to a mean particle spacing of 3.1$\mu$m. An energy gap of $\Delta E \approx 25 E_0 \approx 45\hbar$Hz (half the limiting value) could be realized in the Tonks-Girardeau regime with a barrier rotating at angular speed $\omega=0.29 \times2\pi$Hz.
Such states are much less fragile than the related NOON states
and we expect a robustness of $\bar{Q}^{[-1]}> 1-1/N=0.99$, based on extrapolating from Fig.~\ref{fig:loss}a. 

\section{Acknowledgments}
We acknowledge stimulating discussions with Bill Phillips, Keith Burnett, Kris Helmerson, Mikkel Anderson, and Boris Svistunov. This work was supported by EuroQUASAR and by the Marsden Fund (contract MAU0706), administered by the Royal Society of New Zealand.

\appendix

\section{Effective Hamiltonian for numerical simulations:}
\label{ap:numerics}
For the numerical calculations we perform an exact diagonalization of the Hamiltonian in a truncated Hilbert space. In order to improve the accuracy of the truncated calculation we have developed an effective Hamiltonian that involves rescaling the interaction constant as detailed in the following.
We start with the second quantized form of the Hamiltonian
$H=H_K+H_B+H_I$, where $H_K$, $H_B$, and $H_I$ are the Hamiltonians describing the kinetic energy of the atoms, the barrier, and the interactions between the atoms, respectively. These are given by
\begin{eqnarray}
H_K &=&\int dx\; \hat{\Psi}^{\dag}(x)\frac{\hbar^2}{2M}\left( -i\frac{\partial}{\partial x} - \frac{\Omega}{L}\right)^2 \hat{\Psi}(x) , \nonumber\\
H_B &=& \int dx \; b \; \delta(x) \; \hat{\Psi}^{\dag}(x)\hat{\Psi}(x) , \nonumber\\
H_I &=& \int dx \; \frac{{g}}{2} \; \hat{\Psi}^{\dag}(x)\hat{\Psi}^{\dag}(x)\hat{\Psi}(x)\hat{\Psi}(x),
\label{eq:H2}
\end{eqnarray} 
where $\hat{\Psi}^{\dag}(x)$ and $\hat{\Psi}(x)$ are the Schr\"odinger field operators with standard bosonic commutation relations. We transform into a truncated angular momentum basis with
\begin{eqnarray}
\hat{\Psi}(x) =\frac{1}{\sqrt{L}}\sum_k e^{i2\pi kx/L} \hat{a}_k,
\end{eqnarray} 
where $\hat{a}^{\dag}_k$ and $\hat{a}_k$ create and destroy an atom with angular momentum $k\hbar$, respectively. We introduce the effective Hamiltonian in the truncated basis by $\tilde{H}=\tilde{H}_K+\tilde{H}_B+\tilde{H}_I$ with
\begin{eqnarray}
\tilde{H}_K &=& \sum_{k=-r/2+1}^{r/2} E_0 \left( k-\frac{\Omega}{2\pi} \right)^2 \hat{a}^{\dag}_k \hat{a}_k , \nonumber\\
\tilde{H}_B &=& \frac{b}{L} \sum_{k_1,k_2=-r/2+1}^{r} \hat{a}^{\dag}_{k_1} \hat{a}_{k_2} ,\nonumber\\
\tilde{H}_I &=& \frac{\tilde{g}}{2L} \sum_{k_1,k_2,q=-r/2+1}^{r/2}  \hat{a}^{\dag}_{k_1} \hat{a}^{\dag}_{k_2}  \hat{a}_{k_1-q} \hat{a}_{k_2+q},
\label{eq:H3}
\end{eqnarray} 
which becomes formally exact and identical to $H$ of Eq.~(\ref{eq:H2}) for $r \to \infty$ and $\tilde{g}=g$. Here, $r$ is the number of angular momentum modes used in the simulation, which we choose to be even. Choosing finite $r$ effectively truncates Hilbert space and constitutes an approximation that is expected to converge towards the exact result for large $r$. We have used up to $r=36$ for testing convergence. The results presented in Figs.~\ref{fig:DEvg} and \ref{fig:loss} have been obtained with $r=20$ (except for Fig.~\ref{fig:loss}a). While at this level the ground and excited state energies have already converged to the level of machine precision for small $g$ up to the NOON-state regime ($g\approx 0.1$), we find slow convergence with increasing $r$ in the strongly-interacting Tonks-Girardeau regime, where we can compare with exact results from the Bose-Fermi mapping~\cite{girardeau_60}.  

Using an effective Hamiltonian for our numerical calculations is a way to improve the accuracy significantly. Formally, the effective Hamiltonian is introduced by a transformation of the full Hamiltonian onto the truncated Hilbert space that preserves a subset of the exact eigenvalues~\cite{Suzuki}. Here, we use a convenient approximate form that involves solely a rescaling of the interaction constant $g$ in Eq.~(\ref{eq:H3}), found by considering the simple system of two interacting particles at $b=0$ and $\Omega=0$, which we solve analytically.

We look for the ground state of $H$ for two particles and write the general wave function as
$|\psi\rangle  = \sum_{k_1,k_2=-\infty}^{\infty}  C_{k_1,k_2}   \hat{a}^{\dag}_{k_1} \hat{a}^\dag_{k_2}|\mbox{vac}\rangle$. Substituting this into the Schr\"{o}dinger equation and making use of bosonic symmetry $C_{k_1,k_2}=C_{k_2,k_1}$ we obtain the set of simultaneous equations
\begin{eqnarray}
\left( E  -  E_0 \left(k_1^2 + k_2^2 \right) \right) C_{k_1, k_2} = \frac{g}{L} \sum_{q=-\infty} ^{\infty}  C_{k_1-q, k_2+q} .
\label{eqn:Cs}
\end{eqnarray}
All terms in the Hamiltonian conserve total angular momentum and we can expect the ground state to have zero angular momentum where $k_1=-k_2$. It follows that the right hand side of Eq.~(\ref{eqn:Cs}) is independent of $k_1$.
Using this it is easy to show that $C_{-q,q}\propto \left(E-2E_0q^2\right)^{-1}$ and
\begin{eqnarray}
\frac{L}{g} =  \sum_{q=-\infty} ^{\infty} C_{-q,q} 
=\frac{L}{\tilde{g}} - \frac{L}{g_0} ,
\label{eq:g0}
\end{eqnarray}
where $L/\tilde{g} = \sum_{q=-r/2+1}^{r/2} C_{-q,q}$ is the result that we would obtain from the truncated Hamiltonian $\tilde{H}$ of Eq.~(\ref{eq:H3}).
The term $-L/g_0$ accounts for the sum over the remaining angular momentum modes and thus defines $g_0$.
From rearranging Eq.~(\ref{eq:g0}) we find that both the energy $E$ as well as the coefficients $C_{-q,q}$ obtained from diagonalizing the effective Hamiltonian $\tilde{H}$ agree with the exact results if 
the  interaction strength is rescaled to 
\begin{equation}
\tilde{g}=g/(1+g/g_0),
\end{equation}
where $g_0$ is given by
\begin{eqnarray}
\frac{L}{g_0} =  \frac{2}{E_0 r} + \frac{2E-E_0}{6E_0^2 r^3} + {\cal O}(r^{-5}).
\end{eqnarray}
The terms beyond the first on the right hand side explicitly depend on the energy $E$ of the solution. Due to their scaling 
with the number of single-particle modes $r$, these terms can be neglected when $r^2 E_0 \gg E - E_0$.
For more than two particles the formally exact effective Hamiltonian contains three and more particle interaction terms, which we will ignore here~\cite{Suzuki}. For the calculations in this work, we approximate the effective Hamiltonian by $\tilde{H}$ of Eq.~(\ref{eq:H3}) with $g_0 = rLE_0/2$.

\section{Single particle spectrum:}
\label{ap:single_particle}
The Hamiltonian describing the system of a 1D loop with a rotating barrier is given in Eq.~(\ref{eq:ham1}). For a single atom analytic solutions can be found.
We write the wave function as  $\Psi(x)=\phi(x)e^{i\Omega x/L}$ and substitute this into the Schr\"{o}dinger equation to obtain
\begin{eqnarray}
-\frac{\hbar^2}{2M}\frac{\partial^2}{\partial x^2} \phi(x) + b \delta(x) \phi(x) = \varepsilon \phi(x),
\label{eq:schrd}
\end{eqnarray} 
where the boundary conditions require $\phi(x)=\phi(x+L) e^{i\Omega}$. The first derivative of the wave function at the Dirac delta barrier is discontinuous and is found by integrating the Schr\"{o}dinger equation over the barrier:
\begin{equation}
\left. \frac{d\phi}{dx} \right|_{x=+0}- \left.\frac{d\phi}{dx} e^{i\Omega}\right|_{x=L-0} = \frac{2Mb}{\hbar^2}\phi(0).
\label{eq:bc2}
\end{equation}
With the ansatz $\phi(x)=e^{i2\pi \alpha x/L} + A e^{-i2 \pi \alpha x/L}$ and using the boundary condition for the wave function we find $A=e^{i2\pi\alpha}S$ and $S=\sin(\pi \alpha+\Omega/2) / \sin(\pi \alpha-\Omega/2)$. Substituting this into Eq. (\ref{eq:bc2}) 
yields
\begin{equation}
\frac{4\pi \hbar^2 \alpha}{MLb}=\cot(\pi \alpha-\Omega/2)+\cot(\pi\alpha+\Omega/2),
\label{eq:alpha}
\end{equation}
which has to be solved for $\alpha$. The discrete solutions $\alpha_\mu$ with $\mu = 0, 1, 2, \ldots$ correspond to the single-particle energy levels $\varepsilon_\mu = \alpha_\mu^2 E_0$.
For the case of  $\Omega=\pi$, Eq.~(\ref{eq:alpha}) is simplified to $2\pi\hbar^2 \alpha_\mu / MLb = -\tan(\pi\alpha_\mu)$ for odd $\mu$ and $\alpha_{\mu} = (\mu+1)/2$ for even $\mu$.

\section{Conditions for existence and energy gap for NOON state:}
\label{ap:noon}
The values of the experimental parameters needed to create a NOON state in this system are determined by three factors~\cite{hallwood_07}. Firstly we want the states $|0,N,0\rangle$ and $|0,0,N\rangle$ to be degenerate (here we describe the state of the system in the occupation number basis where the three numbers in the ket represent the $-\hbar$, 0 and $\hbar$ angular momentum modes, respectively). This is achieved when $\Omega=\pi$, which is the condition we are considering here.

Secondly, the two states must be, at most, weakly coupled to and energetically well separated from other states, otherwise the ground state will be a superposition of many states. This is achieved by requiring a minimum interaction strength to energetically separates the states  $|0,N-n,n\rangle$, where $n=1,...,N-1$ from $|0,N,0\rangle$ and $|0,0,N\rangle$. At the same time, the interaction should be weak enough to not couple strongly to other states through interactions. We can estimate these criteria by considering two simple two level systems. First we consider the states $|0,N,0\rangle$ and $|0,N-1,1\rangle$, where $|0,N-1,1\rangle$ is the state that is coupled strongest to $|0,N,0\rangle$ through the barrier term in the Hamiltonian. The Hamiltonian for this two state system is,
\begin{eqnarray}
H_{e} &=& \frac{{g}}{L}(N-1) |0,N-1,1\rangle \langle 0,N-1,1| \nonumber \\
&&+ \frac{b}{L}\sqrt{N} \left( |0,N,0\rangle \langle 0,N-1,1| \right.\nonumber \\
&&\;\;\;\;\;\;\;\;\;\;\;\;\;\;\;\;\;\;\;\;\left.+ |0,N-1,1\rangle \langle0, N,0| \right)
\end{eqnarray} 
where we have ignored a constant energy term. Starting with the ansatz $|\Psi\rangle = a_0|0,N,0\rangle+a_1 |0,N-1,1\rangle$ we find
\begin{eqnarray}
\left|\frac{a_0}{a_1}\right| = \frac{g(N-1)}{2b\sqrt{N}} + \sqrt{\left(\frac{g(N-1)}{2b\sqrt{N}}\right)^2 + 1}.
\end{eqnarray}

We must also consider the relative amplitudes of the states $|0,N,0\rangle$ and $|1,N-2,1\rangle$, where $|1,N-2,1\rangle$ is the state that is coupled strongest to $|0,N,0\rangle$ through the interaction term in the Hamiltonian. The Hamiltonian for this two state system is,
\begin{eqnarray}
H_{e} &=& \left[ 2E_0 + \frac{{g}}{L}(2N-3) \right] |1,N-2,1\rangle \langle 1, N-2,1|  \nonumber \\
&&+ \frac{g}{2L}\sqrt{N(N-1)} \left( |0,N,0\rangle \langle 1, N-2,1| \right. \nonumber \\
&&\;\;\;\;\;\;\;\;\;\;\;\;\;\;\;\;\;\;\;\;\;\;\;\;\;\;\;\;\;\;\left.+ |1,N-2,1\rangle \langle 0,N,0| \right)
\end{eqnarray} 
Again we have ignored a constant energy term. This time we start with the ansatz $|\Psi\rangle = a_0|0,N,0\rangle+\tilde{a}_1 |1,N-2,1\rangle$ and find,
\begin{eqnarray}
\left|\frac{a_0}{\tilde{a}_1}\right| &=& \frac{2E_0L+g(2N-3)}{g\sqrt{N(N-1)}} \nonumber\\
&&+ \sqrt{\left(\frac{(2E_0L+g(2N-3)}{g\sqrt{N(N-1)}}\right)^2 + 1}.
\end{eqnarray}
To create a NOON state we require $|a_0/a_1|,|a_0/\tilde{a}_1| \gg 1$, which is achieved when,
\begin{eqnarray}
\frac{b\sqrt{N}}{L} \ll \frac{gN}{2L} \ll E_0.
\label{eq:condition}
\end{eqnarray}
We see that the mean interaction energy per particle needs to be much smaller than the kinetic energy per particle and much larger than the barrier energy times the square root of the number of particles. $E_0$ is inversely proportional to the square of the circumference of the loop, $L^2$, and the mass of the atoms, $M$. Therefore, it is easier to fulfill condition (\ref{eq:condition}) with lighter atoms and smaller rings. 
As the number of particles are increased it will be experimentally unattainable to reach the small interaction strength and the barrier height needed to make a NOON state.

Finally, in order to obtain the superposition, the states $|0,N,0\rangle$ and $|0,0,N\rangle$ must be coupled. By calculating the coupling strength we will also be able to calculate the energy level splitting between the ground and first excited states. There is no direct first order coupling between the two states, but they do couple through intermediate states. We have already argued that the population in states other than $|0,N,0\rangle$ and $|0,0,N\rangle$ is small. In this regime and at $\Omega = \pi$ it is a good approximation to consider just the 0 and 1 angular momentum modes. States that have atoms with other angular momentum modes have a larger kinetic energy associated with them. This makes the states energetically unfavorable and only provide a small addition to the coupling strength between $|0,N,0\rangle$ and $|0,0,N\rangle$. 

\begin{figure}[th]
\includegraphics[width=8cm]{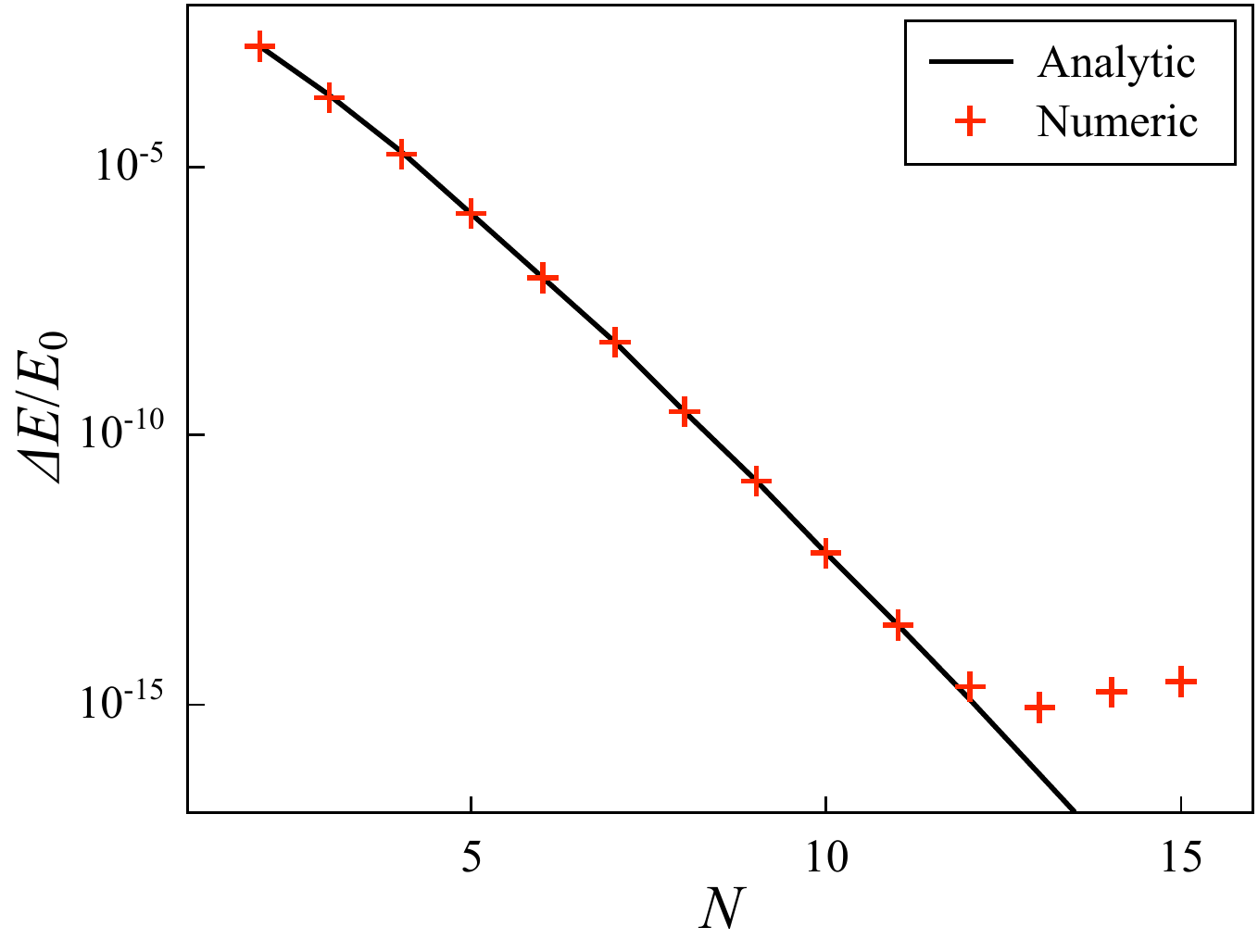}
\caption{(Color online)
NOON state energy level splitting as a function of atom number. The line shows the energy level splitting calculated using Eq.~(\ref{eq:V}),
where $b/L=0.008E_0$ and $g/L=4\pi b\sqrt{N}/L(N-1)$. In the regime where the NOON state is formed it is adequate to describe the system using just the 0 and $\hbar$ angular momentum modes. The results from the numerical simulation are shown by the red crosses. The energy level splitting decreases rapidly as the number of particles are increased. We see the numerical result break down for $N>11$ due to limited numerical accuracy. The analytic result is still valid beyond this point.}
\label{fig:noon}
\end{figure}

The Schr\"{o}dinger equation for this system can be written as a set of simultaneous equations,
\begin{eqnarray}
\lambda a_n = t_n a_n + V_{n,n-1} a_{n-1} + V_{n,n+1}a_{n+1},
\label{eqn:twomode}
\end{eqnarray}
Here $a_n$ is the coefficient of state $|0,N-n,n\rangle$, $\lambda$ are the eigenenergies that solve the system, $t_n = \langle 0,N-n,n|H|0,N-n,n\rangle = gn(N-n)/L + \mbox{const.}$, $V_{n,n+1} = \langle0, N-n,n|H|0,N-n-1,n+1\rangle = b\sqrt{(N-n)(n+1)}/L$ and $H$ is the Hamiltonian. We can systematically eliminate coefficients of intermediate states leaving just $a_0$ and $a_N$,
\begin{equation}
a_N = \left[ \frac{(\lambda-t_0)(\lambda-t_1)...(\lambda-t_{N-1})}{V_{01}V_{12}...V_{N-1,N}} + A_N \right] a_0
\end{equation}
To prove by induction that this is the general form we add another atom to the system and eliminate $a_N$ using the additional equation $\lambda a_N = t_N a_N + V_{N,N-1} a_{N-1} + V_{N,N+1}a_{N+1}$. This leaves just $a_0$ and $a_{N+1}$  and gives
\begin{equation}
a_{N+1} = \left[ \frac{(\lambda-t_0)(\lambda-t_1)...(\lambda-t_{N})}{V_{01}V_{12}...V_{N,N+1}} + A_{N+1} \right] a_0 ,
\end{equation}
where
\begin{eqnarray}
A_{N+1} &=& A_N\frac{(\lambda-t_N)}{V_{N,N+1}}  \nonumber\\
&&- \frac{V_{N,N-1}}{V_{N,N+1}}\frac{(\lambda-t_0)(\lambda-t_1)...(\lambda-t_{N-2})}{V_{01}V_{12}...V_{N-2,N-1}} \nonumber \\
&&- A_{N-1}\frac{V_{N,N-1}}{V_{N,N+1}}
\end{eqnarray}
and $A_1=0$.

We are now left with two simultaneous equations,
\begin{eqnarray}
\lambda a_0 &=& t(\lambda) a_0 + V(\lambda) a_{N} , \nonumber \\
\lambda a_N &=& t(\lambda) a_N + V(\lambda) a_{0} ,
\label{eq:a0aN}
\end{eqnarray}
where
\begin{eqnarray}
V(\lambda) &=& \frac{V_{01} V_{12} ... V_{N-1,N}}{(\lambda-t_1)(\lambda-t_2)...(\lambda-t_{N-1})} , \nonumber \\
t(\lambda) &=&  t_0-A_N V(\lambda) .
\label{eq:num}
\end{eqnarray}
When the amplitude of states other than $|0,N,0\rangle$ and $|0,0,N\rangle$ are small, Eq.~(\ref{eq:a0aN}) gives the ground and first excited state energies, because $|0,N,0\rangle$ and $|0,0,N\rangle$ have the lowest energies of the uncoupled system. If we assume $t(\lambda_0)\approx t(\lambda_1)\approx t(t_0)$ and $V(\lambda_0) \approx V(\lambda_1) \approx V(t_0)$ then we have a simple two level system and the energy level splitting is given by $\Delta E = \lambda_1-\lambda_0 = 2 V$. We obtain Eq.~(\ref{eq:V}).
For large $N$ this expression is dominated by the inverse factorial and thus becomes exceedingly small. We have verified the assumptions by expanding around $t_0$ and find it is valid under the conditions given in Eq.~(\ref{eq:condition}). We compare Eq.~(\ref{eq:V}) with a full numerical solution of the two mode model of Eq.~(\ref{eqn:twomode}) in Fig.~\ref{fig:noon}.


\begin{thebibliography}{99}
\bibitem{note1} NOON states are also called generalized GHZ states after 
Ref.~\cite{greeneberger89}, which discusses entanglement properties of a NOON state of three spin-1/2 particles.
\end{thebibliography}
\end{document}